\documentclass[11 pt,a4paper]{article}
\usepackage[utf8]{inputenc}
\usepackage{amsmath}
\usepackage{amsfonts}
\usepackage{amssymb}
\usepackage{pgfplots}
\usepackage{xcolor}
\usepgfplotslibrary{polar}
\usepackage{tikz}
\usepackage{subcaption}
\usepgfplotslibrary{fillbetween}
\usetikzlibrary{patterns}
\usepackage{indentfirst}
\usepackage{amsmath}

\author{Jacques Balayla\textsuperscript{1}
\footnote{\textbf{Author Contributions:} Jacques Balayla - formal analysis, methodology, writing original draft, writing review and editing. 
\textbf{Competing Interests}: The author has declared that no competing interests exist. \textbf{Financial Disclosure}: The author received no specific funding for this work.
\textbf{\textsuperscript{1}To whom correspondence should be addressed:} Dr. Jacques Balayla MD, MPH, CIP, FRCSC. Quilligan Scholar. Osler Fellow. Department of Obstetrics and Gynecology. McGill University, Montreal, Quebec, Canada. e-mail: jacques.balayla@mcgill.ca }}

\title{Prevalence Threshold ($\phi_e$) and the Geometry of Screening Curves}

\date{}

\begin{document}
\maketitle 
 
\begin{abstract}

The relationship between a screening tests' positive predictive value, $\rho$, and its target prevalence, $\phi$, is proportional - though not linear in all but a special case. In consequence, there is a point of local extrema of curvature defined only as a function of the sensitivity $a$ and specificity $b$ beyond which the rate of change of a test's $\rho$ drops precipitously relative to $\phi$. Herein, we show the mathematical model exploring this phenomenon and define the $prevalence$ $threshold$ ($\phi_e$) point where this change occurs as: 

\begin{center}
\begin{Large}
$\phi_e=\frac{\sqrt{a\left(-b+1\right)}+b-1}{(\varepsilon-1)} = \frac{\sqrt{a\left(-b+1\right)}+b-1}{J} = \frac{\sqrt{1-b}}{\sqrt{a}+\sqrt{1-b}}$ 
\end{Large}
\end{center}
\
where $\varepsilon$ = $a$+$b$ and $J$ = $a$+$b$-1 (Youden's $J$ statistic). From the prevalence threshold we deduce a more generalized relationship between prevalence and positive predictive value as a function of $\varepsilon$, which represents a fundamental theorem of screening, herein defined as: 
\\
\
\begin{center}
\begin{large}
$\displaystyle\lim_{\varepsilon \to 2}{\displaystyle \int_{0}^{1}}{\rho(\phi)d\phi} = 1$
\end{large}
\end{center}
\

Understanding the concepts described in this work can help contextualize the validity of screening tests in real time, and help guide the interpretation of different clinical scenarios in which screening is undertaken. 
\end{abstract} 
\newpage

\section{Introduction}
 
Screening is defined as the presumptive identification of unrecognised disease in asymptomatic individuals by means of tests, examinations or procedures \cite{wilson1968principles}. The ultimate purpose of a screening test is two-fold: 1) to allow for the early detection of a disease, and thus establish a surveillance plan to assess progression, and/or 2) to detect a condition early in order to treat it most effectively. Screening tests are not considered diagnostic, but are used to identify a subset of the population that should undergo additional testing in order to accurately establish the presence or absence of disease \cite{sackett1975screening}.
\\
\

In 1968, the World Health Organization (WHO) published guidelines on the principles and practice of screening for disease, which are often referred to as the $Wilson-Jungner$ criteria \cite{andermann2008public}. These principles are still broadly applicable today and include the following: 1) The condition should be an important health problem.
2) There should be a treatment for the condition.
3) Facilities for diagnosis and treatment should be available.
4) There should be a latent stage of the disease.
5) There should be a screening test or examination for the condition.
6) The test should be acceptable to the population.
7) The natural history of the disease should be adequately understood.
8) There should be an agreed policy on whom and when to treat.
9) The total cost of finding a case should be economically balanced in relation to medical expenditure as a whole.
Finally, 10) Case-finding should be a continuous process.
\\
\

In keeping with these ideas, it is important to contextualize them into the natural disease process (Figure 1). The biological onset of disease is followed by clinical symptoms, then diagnosis and therapy until there is an outcome, including survival or death \cite{herman2006makes}. The time from the onset of disease until clinical symptoms occur is known as the pre-clinical phase. The individual has the disease but doesn't know it. The clinical phase is the latter part of the process, from the occurrence of clinical symptoms through therapy and outcome \cite{herman2006makes}. Within the preclinical phase, there may be an interval between the onset of the disease and the occurrence of clinical symptoms during which disease can be detected with certain screening tests. This is called a detectable pre-clinical, or latent, phase. If treatment is more effective during the preclinical stage of disease, as is the case for most conditions, screening for disease during the detectable pre-clinical phase offers an advantage \cite{herman2006makes}.
\\
\
\begin{figure}[ht!]
\centering
\includegraphics[scale=1.05]{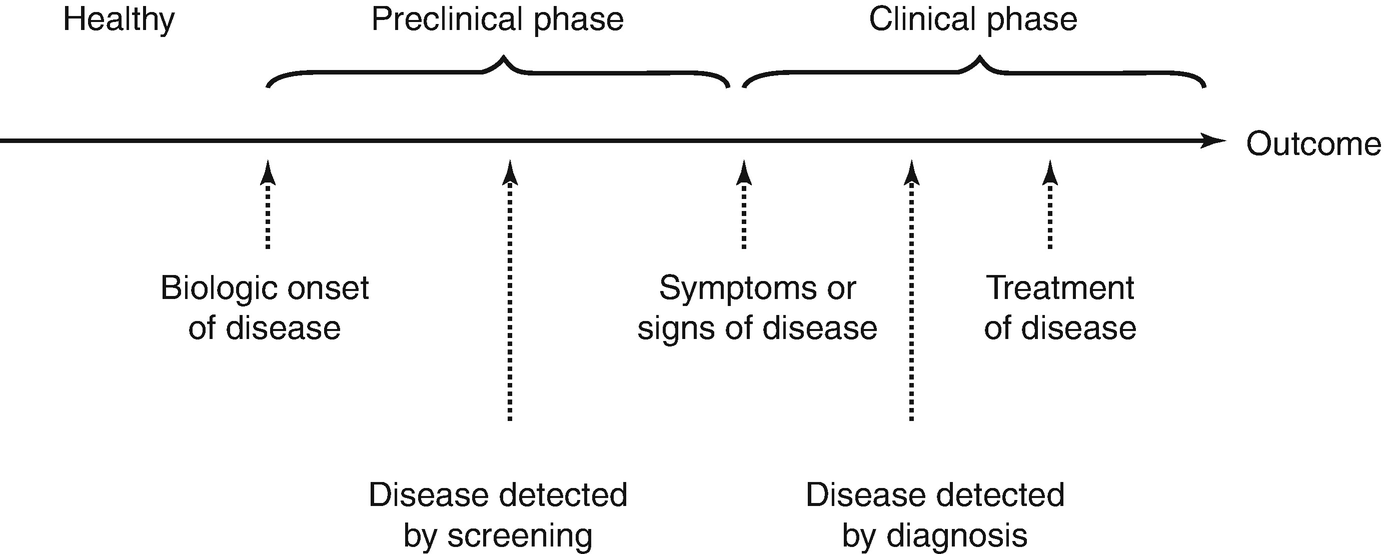}
\\
\

\textbf{Figure 1}. Natural Progression of Disease
\end{figure}
\\
\newpage
When conducting a screening test, 4 different parameters help to determine its overall ability to correctly identify individuals with the disease in question \cite{brenner1997variation}. These include the sensitivity $a$, specificity $b$, positive predictive value $\rho$ and negative predictive value $\sigma$. Sensitivity refers to the proportion of affected individuals that have a positive test (true positive rate), and specificity refers to the proportion of unaffected individuals that have a negative test (true negative rate). On the other hand, the positive predictive value ($\rho$) is defined as the percentage of patients with a positive test that do in fact have the disease, and conversely, the negative predictive value ($\phi$) refers to the percentage of patients with a negative test that do not have the disease. To further explore these properties in detail, we draw a 2 x 2 confusion matrix  (Table 1) as follows:
\\
\

\textbf{Table 1}. 2x2 Confusion Matrix
\begin{table}[h]
\centering
\begin{tabular}{ccc}
                                             & \multicolumn{2}{c}{\textbf{Condition}}                                              \\ \cline{2-3} 
\multicolumn{1}{c|}{}                        & \multicolumn{1}{c|}{\textbf{Present}}    & \multicolumn{1}{c|}{\textbf{Absent}}     \\ \hline
\multicolumn{1}{|c|}{\textbf{Positive Test}} & \multicolumn{1}{c|}{True Positive ($\alpha$)}  & \multicolumn{1}{c|}{False Positive ($\beta$)} \\ \hline
\multicolumn{1}{|c|}{\textbf{Negative Test}} & \multicolumn{1}{c|}{False Negative ($\gamma$)} & \multicolumn{1}{c|}{True Negative ($\delta$)}  \\ \hline
\end{tabular}
\end{table}

Where the following variables are thus defined:
\begin{center}
prevalence = $\phi = {(\alpha + \gamma})/{(\alpha+\beta+\gamma+\delta})$,
\

sensitivity = $a ={\alpha}/{(\alpha+\gamma)}$,
\ 

specificity = $b ={\delta}/{(\delta+\beta)}$, 
\

PPV = $\rho = {\alpha}/{(\alpha+\beta)}$, 
\

NPV = $\sigma = {\delta}/{(\gamma+\delta)}$. 
\end{center}

\section{Bayes’ Theorem}
Bayes’ Theorem describes the probability of an event based on prior knowledge of conditions  related to that specific event \cite{moons1997limitations}. Mathematically speaking, the equation translates to the conditional probability of an event A given the presence of an event or state B. As per Bayes' Theorem, the above relationship is equal to the probability of event B given event A, multiplied by the ratio of independent probabilities of event A to event B. Simply stated, the equation is written as follows:

\begin{large}
\begin{equation}
P(A|B) = \frac{P(B|A) P(A)}{P(B)}
\end{equation}
\end{large}

Where A, B = events, $P(A|B)$ = probability of A given B is true, $P(B|A)$ = probability of B given A is true, and P(A) and P(B) are the independent probabilities of A and B. If we use T +/- as either a positive or negative test, and denote D +/- as the presence or absence or disease then we can use Bayes' theorem to calculate the positive predictive value by asking the following question: given a positive screening test, what is the probability that an individual does in fact have the disease in question?

\begin{large}
\begin{equation}
P(D+|T+)= \frac{P(T+|D+)P(D+)}{P(T+|D+)P(D+)+P(T+|D-)P(D-)}
\end{equation}
\end{large} 

Since the probability of not having the disease is equal to the complement of the prevalence and the false positive rate is equal to the complement of the specificity, Bayes' theorem yields the PPV as follows:

\begin{large}
\begin{equation}
\rho(\phi) = \frac{a\phi}{ a\phi+(1-b)(1-\phi)} =\frac{a\phi}{a\phi+ b\phi - b - \phi +1}
\end{equation}
\end{large} 
\\
\
where $\rho(\phi)$ = PPV, a = sensitivity, b = specificity and $\phi$ = prevalence.
\\
\

We have thus shown that the PPV, $\rho$, is a function of prevalence, $\phi$. As the prevalence increases, the $\rho(\phi)$ also increases but the NPV, $\sigma(\phi)$, decreases and vice-versa.
\newpage

By the above equation, we obtain:

\begin{large}
\begin{equation}
\lim_{\phi \to 1} \rho(\phi) = 1
\end{equation}
\end{large}

and,

\begin{large}
\begin{equation}
\lim_{\phi \to 0} \rho(\phi) = 0
\end{equation}
\end{large}

Inversely, we can isolate the prevalence as a function of sensitivity, specificity and the PPV as follows:

\begin{large}
\begin{equation}
\phi = \frac{1 - b}{\frac{a}{\rho}-a-b+1} 
\end{equation}
\end{large} 

It is important to bear in mind that screening curves come in two forms: one prevalence-independent relating the sensitivity to the specificity, also known as the receiver operating characteristc (ROC) curve, and one prevalence-dependent relating a tests' positive predictive value to its target disease's prevalence - as depicted in this work \cite{fawcett2006introduction}. The latter screening curves are continuous, positive functions in the real plane, whose domain spans $0<\phi<1$ and cross the spectrum boundaries at coordinates [0,0] and [1,1]. The relationship between $\phi$ and $\rho$ is proportional and as such, these curves retain their concavity or convexity throughout the domain.  

\section{The Screening Paradox}
If a disease process is recognized and treated early, and a diagnosis is therefore prevented, the prevalence of such disease would drop in the population, which as per Bayes' theorem, would make the tests’ predictive value drop in return \cite{smith2000first}. Put another way, assuming as per $Wilson-Jungner$ criteria that a curative/preventative treatment following an abnormal screening test exists, a very powerful screening test would, by performing and succeeding at the very task it was developed to do, paradoxically reduce its ability to correctly identify individuals with the disease it screens for in the future. Now, this paradoxical effect tends to be well tolerated by the system up to a well defined prevalence point beyond which the geometry of the screening curve changes most drastically. Technically speaking, there is a prevalence level below which the rate of change of a test's $\rho$ drops precipitously relative to $\phi$. In order to explore this notion further, we define a new entity henceforth known as the screening coefficient, $\varepsilon$, defined as the sum of the sensitivity and specificity, a + b.

\section{The Screening Coefficient ($\varepsilon$)}
To preface this section, we hereby define a new entity, the screening coefficient ($\varepsilon$), as the sum of sensitivity $a$ and specificity $b$, which is related to Youden's $J$ statistic.
\\
\
\begin{large}
\begin{equation}
\varepsilon = a + b  \rightarrow [\varepsilon\in {\rm I\!R} | 0 < \varepsilon < 2] \rightarrow J = \varepsilon - 1
\end{equation}
\end{large} 
\\
\
We know from equation (3) that an increase in prevalence will bring about an increase in the PPV (and vice-versa) at different velocities depending on the prevalence/pre-test probability level. We can calculate this velocity by taking the first order derivative of equation (3) as follows:
\\
\
\begin{large}
\begin{equation}
\frac{d\rho}{d\phi} = \frac{a(1 - b)}{(a\phi+(1-b)(1-\phi))^2} 
\end{equation}
\end{large} 
\\
\
Since both $\phi$ and $\rho$ are positive real numbers between 0 and 1, d$\rho$/d$\phi$ is a positive real number as well as per equation (8). This implies that the relationship between $\phi$ and $\rho$ is directly proportional throughout the interval [0 - 1] $\in$ ${\rm I\!R}$. However, in order to determine whether the rate at which the PPV is  changing with respect to prevalence is accelerating or decelerating, we take the second order derivative of equation (3) as follows:
\\
\
 \begin{large}
\begin{equation}
\frac{d^2\rho}{d\phi^2} = -\frac{2a\left(-b+1\right)\left(a-1+b\right)}{\left(a\phi+\left(1-b\right)\left(1-\phi\right)\right)^3}
\end{equation}
\end{large} 
\\
\

From equation (9) it follows that when:
\\
\
\begin{large}
\begin{equation}
\varepsilon < 1\Rightarrow \frac{d^2\rho}{d\phi^2} > 0 
\end{equation}
\end{large} 
\\
\
\begin{large}
\begin{equation}
\varepsilon > 1 \Rightarrow \frac{d^2\rho}{d\phi^2} < 0 
\end{equation}
\end{large}
\\
\
\begin{large}
\begin{equation}
\varepsilon = 1 \Rightarrow \frac{d^2\rho}{d\phi^2} = 0
\end{equation}
\end{large}
\\
\

In order to illustrate the above concepts, let us define a hypothetical condition. Condition X is a disease present in a population. It has a preclinical phase and is amenable to screening. Test X is the screening test developed to detect the latent phase of Condition X. Test X therefore has all of the pertinent screening parameters - sensitivity, specificity, and negative and positive predictive values. If condition X has a high prevalence in the population (e.g. hyperlipidemia, hypertension, diabetes, endemic infections, amongst others) or a high pre-test probability in a given individual and $\varepsilon >1$, then significant drops in prevalence will not bring about significant drops in PPV until prevalence drops below a certain threshold, which for cases of $\varepsilon >1$, occurs at low prevalence levels. It thus follows that in cases like this, the screening tests detection ability remains relatively stable until it has significantly helped drop the prevalence. On the other hand, if condition X has a high prevalence in the population and $\varepsilon<1$, then small drops in prevalence will bring about significant drops in PPV until prevalence drops below a certain threshold at a higher prevalence (Figure 2).
\\
\ 
\begin{center}
\begin{tikzpicture}
	\begin{axis}[
    axis lines = left,
    xlabel = $\phi$,
	ylabel = {$\rho(\phi)$},    
     ymin=0, ymax=1,
    legend pos = outer north east,
     ymajorgrids=false,
    grid style=dashed,
    width=10cm,
    height=5cm,
     ]
	\addplot [
	domain= 0:1,
	color= blue,
	]
	{(0.60*x)/((0.60*x+(1-0.95)*(1-x))};

\addplot [
dashed,	
	domain= 0:1,
	color= blue,
	]
	{0.40*(-0.95+1)/((.40*x)+(1-0.95)*(1-x))^2};

\addplot [	
domain=0.070:0.32, samples=100, color=red] {(1.4*x+0.475)};

\addplot [dashed,	
domain=0.0:0.42, samples=100, color=red] {(1.4*x+0.475)};

\addplot [	
domain=0.195:0.235, samples=100, color=red] {(-3.8*x+1.49)};

\addplot+ [
dashed,
domain= 0:1,	
color = black,
mark size = 0pt
 ]
	{1};
	\fill [red] (235,60) circle[radius=2pt];

	\node[above,red] at (60,85) {$\varepsilon>$1};
	\node[above,black] at (240,39) {$\kappa=1/R$};
	\node[above,black] at (120,70) {\scriptsize{$tangent$}};
	\node[above,black] at (235,63) {\scriptsize{$R$}};
	\addlegendentry{$\rho(\phi)$}
	\addlegendentry{$d\rho/d\phi$}
	\addlegendentry{$\rho=1$}
	\end{axis}

\end{tikzpicture}
\end{center}
\begin{center}
\begin{tikzpicture}
 
	\begin{axis}[
    axis lines = left,
    xlabel = $\phi$,
	ylabel = {$\rho(\phi)$},    
     ymin=0, ymax=1,
    legend pos = outer north east,
     ymajorgrids=false,
    grid style=dashed,
    width=10cm,
    height=5cm,
     ]
	\addplot [
	domain= 0:1,
	color= blue,
	]
	{(0.20*x)/((0.20*x+(1-0.40)*(1-x))};

\addplot [
dashed,	
	domain= 0:1,
	color= blue,
	]
	{0.20*(-0.40+1)/((.20*x)+(1-0.40)*(1-x))^2};

\addplot+ [
dashed,
domain= 0:1,	
color = black,
mark size = 0pt
 ]
	{1};
	\node[above,red] at (60,85) {$\varepsilon<$1};
	\addlegendentry{$\rho(\phi)$}
	\addlegendentry{$d\rho/d\phi$}
	\addlegendentry{$\rho=1$}
	\end{axis}

\end{tikzpicture}
\end{center}

\begin{center} \textbf{Figure 2}. The first graph represents  scenarios where $\varepsilon$ $>$ 1. We denote the line tangent to the point of maximum curvature $\kappa$ from which we derive the radius of curvature $R$, perpendicular to it. The second graph represents the more rare scenarios where $\varepsilon$ $<$ 1. The sensitivity and specificity are constant and were randomly chosen to satisfy the $\varepsilon$ condition.
\end{center}

\section{Derivation of the radius of curvature of $\rho(\phi)$}
In order to determine the radius of curvature of the $\rho(\phi)$ graph at any given point M, we consider a circle with radius $R$, which is perpendicular to the tangent line of the function at that point. We consider an adjacent point increment by d$\phi$ and draw another tangent line to this point N, which we join to the center of the circle with radius $R$. As such, an arc of length dS is formed, which in turn creates an angle $\varphi$ between M and N. These variables see the following properties:
\begin{large}
\begin{equation}
tan(\varphi) = \frac{d\rho}{d\phi}
\end{equation}
\end{large}

\begin{large}
\begin{equation}
dS = R{d\varphi} = \sqrt{1 + (\frac{d\rho}{d\phi})^2}d\phi
\end{equation}
\end{large}
\\
\

From equalities (13) and (14), the differential equation follows:

\begin{large}
\begin{equation}
\frac{d}{d\phi}tan(\varphi) = \frac{d}{d\phi}(\frac{d\rho}{d\phi}) = \frac{d^2\rho}{d\phi^2}
\end{equation}
\end{large}

From the trigonometric identity $1 + tan^2(\varphi) = sec^2(\varphi)$, it follows that:
\begin{large}
\begin{equation}
\frac{d}{d\phi}tan(\varphi) = sec^2(\varphi)\frac{d\varphi}{d\phi} = \frac{d^2\rho}{d\phi^2}
\end{equation}
\end{large}

Therefore,

\begin{large}
\begin{equation}
(1 + tan^2(\varphi))\frac{d\varphi}{d\phi} = \frac{d^2\rho}{d\phi^2}
\end{equation}
\end{large}

Since $tan(\varphi) = d\rho/d\phi$, equation (17) becomes:

\begin{large}
\begin{equation}
(1 + (\frac{d\rho}{d\phi})^2)\frac{d\varphi}{d\phi} = \frac{d^2\rho}{d\phi^2}
\end{equation}
\end{large}

Isolating $d\varphi/d\phi$, we obtain:

\begin{large}
\begin{equation}
\frac{d\varphi}{d\phi} = \frac{\frac{d^2\rho}{d\phi^2}}{(1 + (\frac{d\rho}{d\phi})^2)}
\end{equation}
\end{large}

Using equation (14) this relationship then becomes:

\begin{large}
\begin{equation}
R\frac{\frac{d^2\rho}{d\phi^2}}{(1 + (\frac{d\rho}{d\phi})^2)} = \sqrt{1 + (\frac{d\rho}{d\phi})^2}
\end{equation}
\end{large}

Finally, isolating the radius of curvature $R$:

\begin{large}
\begin{equation}
R= \frac{[1 + (\frac{d\rho}{d\phi})^2]^\frac{3}{2}}{|\frac{d^2\rho}{d\phi^2}|}
\end{equation}
\end{large}

The radius of curvature $R$ is inversely proportional to $\kappa$ such that:
\begin{large}
\begin{equation}
R = \frac{1}{\kappa} \Rightarrow \kappa = \frac{|\frac{d^2\rho}{d\phi^2}|}{[1 + (\frac{d\rho}{d\phi})^2]^\frac{3}{2}}
\end{equation}
\end{large}

Now that we know what the curvature function $\kappa$ is, we can determine where the curvature of $\phi(\rho)$ falls at a maximum. Practically speaking, this represents the point of sharpest change in $\frac{d\rho}{d\phi}$, known as the extrema. 
In order to do so, we find the derivative of the $\kappa$ function and determine its roots:
\begin{large}
\begin{equation}
\frac{d\kappa}{d\phi}=0 \hookrightarrow\lbrace\phi_e,\rho_e\rbrace 
\end{equation}
\end{large}

The above equation yields the value of $\phi$ where the maximum curvature $\kappa$ and thus a minimum radius of curvature $R$ exist. We define this point as the point of local extrema [$\phi_e,\rho_e$] of the $\rho(\phi)$ function. On the other hand, the inflection point [$\phi_i,\rho_i$] is a point on a curve at which the sign of the curvature (i.e., the concavity) changes. The points of local extrema are distinguishable from the inflection point only in that the curvature function's second order-derivative equals 0:

\begin{large}
\begin{equation}
\frac{d^2\kappa}{d\phi^2}=0 \hookrightarrow\lbrace\phi_i,\rho_i\rbrace 
\end{equation}
\end{large}

However, as we described previously, given the proportionality between $\phi$ and $\rho$ all screening curves retain their concavity/convexity throughout the domain [0,1] as a function of a and b, and thus no inflection points are observed in these curves. Conversely, the point of local extrema $\phi_e,\rho_e$ tells us where the sharpest turn, or change, in PPV as a function of prevalence occurs. In cases of when $\varepsilon >1$  the sharp increase occurs at lower prevalence levels with higher PPV levels, and vice-versa for $\varepsilon <1$.
\\
\
\newpage
By equating equation (22) to 0 and looking for its roots, we re-arrange the terms and the above expression simplifies to:

\begin{large}
\begin{equation}
1 = -\frac{a^{2}\left(-b+1\right)^{2}+\left(a\phi+\left(-b+1\right)\left(-\phi+1\right)\right)^{4}}{2\left(a\phi+\left(-b+1\right)\left(-\phi+1\right)\right)^{4}}
\end{equation}
\end{large}

\begin{large}
\begin{equation}
\left(a\phi+\left(-b+1\right)\left(-\phi+1\right)\right)^{4} = -a^{2}\left(-b+1\right)^{2}
\end{equation}
\end{large}

Taking the fourth root of both sides, we obtain:

\begin{large}
\begin{equation}
\left(a\phi+b\phi-b-\phi+1\right) = \pm\sqrt{a\left(-b+1\right)}
\end{equation}
\end{large}

Expanding and isolating $\phi$ while taking the positive value of the root so that the value obtained may fall inside the domain of the function, we obtain:
\begin{large}
\begin{equation}
\phi_e = \frac{\sqrt{a\left(-b+1\right)}+b-1}{(a+b-1)}=\frac{\sqrt{a\left(-b+1\right)}+b-1}{(\varepsilon-1)}
\end{equation}
\end{large}

This is the value of prevalence where the point of local extrema $\phi_e$ of $\rho(\phi)$ is found. We denote this value of $\phi$ as the \textit{prevalence threshold}. By plugging $\phi_e$ into equation (3) we obtain its corresponding $\rho_e$ value. Note the inverse relationship between $\phi_e$ and $\varepsilon$ and thus with the Youden's $J$ statistic.

\begin{large}
\begin{equation}
\phi_e \sim \frac{1}{\varepsilon}\sim \frac{1}{J}
\end{equation}
\end{large}

It is critical to understand that an identical value of 
$\varepsilon$ can provide significantly different prevalence thresholds as sensitivities and specificities do not respect commutative laws in this context. Since the specificity is a measure of the true negative rate, slight changes in specificity provide greater changes in the positive predictive value. In keeping with this idea, the equation for the prevalence threshold contains the specificity parameter $b$ thrice whereas the sensitivity parameter $a$ appears only twice, indeed implying the prevalence threshold is more sensitive to changes in specificity, even for a fixed $\varepsilon$. For a given $\varepsilon$, the higher the specificity, the lower the prevalence threshold and the sharpest the curvature of the local extrema. 
\newpage
Using radical conjugates, we can further simplify the prevalence threshold equation into its most basic form - without the need for the Youden's $J$ statistic statistic:
\subsection{Using radical conjugates to simplify $\phi_e$}
As stated, we can use radical conjugates to further simplify the prevalence threshold equation. Let c = 1-b, the complement of the specificity, otherwise known as the fall-out or false positive rate (FPR). The $\phi_e$ equation thus becomes:

\begin{large}
\begin{equation}
\phi_e = \frac{\sqrt{ac}-c}{a-c}
\end{equation}
\end{large}

Multiplying by its radical conjugate, we obtain:

\begin{large}
\begin{equation}
\phi_e = \frac{\sqrt{ac}-c}{a-c} \left[\frac{\sqrt{ac}+c}{\sqrt{ac}+c}\right]
\end{equation}
\end{large}

The square difference in the numerator yields:

\begin{large}
\begin{equation}
\phi_e = \frac{ac-c^2}{a\sqrt{ac}+ac-c\sqrt{ac}-c^2} 
\end{equation}
\end{large}

Factoring out c, and knowing that $\frac{\sqrt{x}}{x}$ is equal to $\frac{1}{\sqrt{x}}$we obtain:

\begin{large}
\begin{equation}
\phi_e = \frac{a-c}{{\frac{a}{c}\sqrt{ac}}+a-\sqrt{ac}-c} = \frac{a-c}{{a\frac{\sqrt{a}}{\sqrt{c}}-\sqrt{ac}}+a-c}
\end{equation}
\end{large}

Factoring out $\sqrt{ac}$ in the denominator's first terms leads to:

\begin{large}
\begin{equation}
\phi_e =  \frac{a-c}{{\sqrt{ac}(\frac{a}{c}-1)}+a-c}
\end{equation}
\end{large}

Finally, replacing 1 by c/c and factoring out the ensuing a-c term, we obtain:
\begin{large}
\begin{equation}
\phi_e =  \left[\frac{a-c}{a-c}\right]\frac{1}{\frac{\sqrt{a}}{\sqrt{c}}+1}
\end{equation}
\end{large}
And thus, replacing c by 1-b the simplified version of the equation follows:
\begin{large}
\begin{equation}
\phi_e = \frac{\sqrt{1-b}}{\sqrt{a}+\sqrt{1-b}}
\end{equation}
\end{large}
\newpage
Using the prevalence threshold as a prevalence value, we can calculate the corresponding positive predictive value by plotting $\phi_e$ into the positive predictive value equation to retrieve [$\rho(\phi_e)$,$\phi_e$]. In so doing we obtain:

\begin{large}
\begin{equation}
\rho(\phi_e)=\sqrt{\frac{a}{1-b}}\left[\frac{\sqrt{1-b}}{\sqrt{a}+\sqrt{1-b}}\right]
\end{equation}
\end{large}

Interestingly, the above expression leads to the well known formulation for the positive predictive value as  a function of prevalence and the positive likelihood ratio (LR+), defined as the sensitivity over the compliment of the specificity. 

\begin{large}
\begin{equation}
\rho(\phi_e)=\phi_e\sqrt{\frac{a}{1-b}}
\end{equation}
\end{large}

\subsection{The Fundamental Theorem of Screening} 
While the curvature $\kappa$ and the point of local extrema $\phi_e,\rho_e$ provide a quantitative determination of the prevalence threshold, we can establish a qualitative determination of this statistic as well, which is far more intuitive. We can calculate the area under the curve (AUC) of $\rho(\phi)$ by integrating through the function's domain between [0,0] and [1,1]. Intuitively, the greater the area, the greater $\varepsilon$ must be and vice-versa. From the indefinite integral:

\begin{large}
\begin{equation}
\int{\rho(\phi)d\phi} = \dfrac{a\left(\left(b-1\right)\ln\left(\left|\left(b+a-1\right)\phi-b+1\right|\right)+\left(b+a-1\right)\phi\right)}{\left(a+b-1\right)^2}
\end{equation}
\end{large}

It thus follows that:

\begin{large}
\begin{equation}
\lim_{\varepsilon \to 2}\int_{0}^{1}{\rho(\phi)d\phi} = 1
\end{equation}
\end{large}

We deduce that as $\varepsilon$ approaches its maximum possible value of 2, the AUC of $\phi(\rho)$ goes to 1. As equation (33) describes the relationship between all of the pertinent parameters of a positive screening test as a function of prevalence, we define the latter as a \textit{fundamental theorem of screening}. Since we know from equation (29) that the $\phi_e$ is inversely proportional to $\varepsilon$, we infer that the greater the AUC, the lower the prevalence threshold and vice-versa. 
\\
\

\subsection{Clinical Corollaries}

All screening parameters are fundamental to the understanding of the value of screening tests, their limitations, and the concepts thus far described in this work. That said, we can consider the predictive values $\rho(\phi)$ to be most consequential to the individual clinician over the other parameters. Why is $\rho(\phi)$ a more critical parameter for the clinician than sensitivity and specificity?  This is simply because the interpretation of predictive values is done at the level of a single test result, among individuals in whom a diagnosis has not yet been made, and whose ultimate diagnostic status is therefore unknown. In the case of sensitivity and specificity, the ultimate diagnostic status in the patient must be known a priori in order to determine whether a particular screening test is sensitive and/or specific. As such, chronologically speaking, since screening tests lead to eventual diagnoses, the interpretation of a test must occur before a diagnosis is made.
\\
\

Figure 3 depicts a random sample of combinations of $\varepsilon$ values calculated from random sensitivities and specificities, and their corresponding prevalence threshold ($\phi_e$) values. To obtain the prevalence at which the threshold is crossed, multiply the values in red by 10. $\phi_e$ is undefined in the special case where the geometry of the screening curve becomes linear as a consequence of $\varepsilon$ equalling 1. Though there is indeed little clinical applicability for tests whose $\varepsilon$ value is $<$ 1, the point of demonstrating the aforementioned cases is to complete the theory for all possible values of prevalence and sensitivity/specificity even if they're not commonly encountered in clinical practice. The reason is simple – sometimes those tests are all that exist for certain conditions. One can contemplate a test whose specificity is poor but whose sensitivity to rule out disease is good so that $\varepsilon$ $\sim$ 1.
\newpage
\begin{center}
\begin{large}
\textbf{Figure 3. Sample screening curves as a function of $\varepsilon$}
\end{large}
\end{center}
\begin{tikzpicture}[trim left=0.3cm]
 
	\begin{axis}[
    axis lines = left,
    xlabel = $\phi$,
	ylabel = {$\rho(\phi)$},    
     ymin=0, ymax=1,
    legend pos = south east,
     ymajorgrids=true,
    xmajorgrids=true,
    grid style=dashed,
    width=5cm,
    height=5cm,
     ]

\addplot [
dashed,	
	domain= 0:1,
	color= blue,
	]
	{(1.0*x)/((1*x+(1-1)*(1-x))};
	\node[above,red] at (700,20) {$\phi_e$ = 0.000}; 
	\addlegendentry{$\varepsilon = 2.00$}
\end{axis}
\end{tikzpicture}%
~%
	\begin{tikzpicture}
 
	\begin{axis}[
    axis lines = left,
    xlabel = $\phi$,
	ylabel = {$\rho(\phi)$},    
     ymin=0, ymax=1,
    legend pos = south east,
     ymajorgrids=true,
    xmajorgrids=true,
    grid style=dashed,
    width=5cm,
    height=5cm,
     ]

\addplot [
dashed,	
	domain= 0:1,
	color= blue,
	]
	{(0.98*x)/((0.98*x+(1-0.97)*(1-x))}; 
	\node[above,red] at (700,20) {$\phi_e$ = 0.148};
	\addlegendentry{$\varepsilon = 1.95$}
\end{axis}
\end{tikzpicture}%
~%
	\begin{tikzpicture}
 
	\begin{axis}[
    axis lines = left,
    xlabel = $\phi$,
	ylabel = {$\rho(\phi)$},    
     ymin=0, ymax=1,
    legend pos = south east,
     ymajorgrids=true,
    xmajorgrids=true,
    grid style=dashed,
    width=5cm,
    height=5cm,
     ]

\addplot [
dashed,	
	domain= 0:1,
	color= blue,
	]
	{(0.95*x)/((0.95*x+(1-0.95)*(1-x))}; 
	\node[above,red] at (700,20) {$\phi_e$ = 0.186};
	\addlegendentry{$\varepsilon = 1.90$}
\end{axis}
\end{tikzpicture}

\begin{tikzpicture}[trim left=0.8cm]
 
	\begin{axis}[
    axis lines = left,
    xlabel = $\phi$,
	ylabel = {$\rho(\phi)$},    
     ymin=0, ymax=1,
    legend pos = south east,
     ymajorgrids=true,
    xmajorgrids=true,
    grid style=dashed,
    width=5cm,
    height=5cm,
     ]

\addplot [
dashed,	
	domain= 0:1,
	color= blue,
	]
	{(0.85*x)/((0.85*x+(1-0.90)*(1-x))}; 
	\node[above,red] at (700,20) {$\phi_e$ = 0.255};
	\addlegendentry{$\varepsilon = 1.75$}
\end{axis}
\end{tikzpicture}%
~%
	\begin{tikzpicture}
 
	\begin{axis}[
    axis lines = left,
    xlabel = $\phi$,
	ylabel = {$\rho(\phi)$},    
     ymin=0, ymax=1,
    legend pos = south east,
     ymajorgrids=true,
    xmajorgrids=true,
    grid style=dashed,
    width=5cm,
    height=5cm,
     ]

\addplot [
dashed,	
	domain= 0:1,
	color= blue,
	]
	{(0.75*x)/((0.75*x+(1-0.85)*(1-x))}; 
	\node[above,red] at (700,20) {$\phi_e$ = 0.309};
	\addlegendentry{$\varepsilon = 1.60$}
\end{axis}
\end{tikzpicture}%
~%
	\begin{tikzpicture}
 
	\begin{axis}[
    axis lines = left,
    xlabel = $\phi$,
	ylabel = {$\rho(\phi)$},    
     ymin=0, ymax=1,
    legend pos = south east,
     ymajorgrids=true,
    xmajorgrids=true,
    grid style=dashed,
    width=5cm,
    height=5cm,
     ]

\addplot [
dashed,	
	domain= 0:1,
	color= blue,
	]
	{(0.65*x)/((0.65*x+(1-0.75)*(1-x))}; 
	\node[above,red] at (700,20) {$\phi_e$ = 0.382};
	\addlegendentry{$\varepsilon = 1.40$}
\end{axis}
\end{tikzpicture}

\begin{tikzpicture}[trim left=0.8cm]
 
	\begin{axis}[
    axis lines = left,
    xlabel = $\phi$,
	ylabel = {$\rho(\phi)$},    
     ymin=0, ymax=1,
    legend pos = north west,
     ymajorgrids=true,
    xmajorgrids=true,
    grid style=dashed,
    width=5cm,
    height=5cm,
     ]

\addplot [
dashed,	
	domain= 0:1,
	color= blue,
	]
	{(0.5*x)/((0.50*x+(1-0.50)*(1-x))}; 
	\node[above,red] at (600,-2) {$\phi_e$ = undefined};
	\addlegendentry{$\varepsilon = 1.00$}
\end{axis}
\end{tikzpicture}%
~%
	\begin{tikzpicture}
 
	\begin{axis}[
    axis lines = left,
    xlabel = $\phi$,
	ylabel = {$\rho(\phi)$},    
     ymin=0, ymax=1,
    legend pos = north west,
     ymajorgrids=true,
    xmajorgrids=true,
    grid style=dashed,
    width=5cm,
    height=5cm,
     ]

\addplot [
dashed,	
	domain= 0:1,
	color= blue,
	]
	{(0.20*x)/((0.20*x+(1-0.2)*(1-x))}; 
	\node[above,red] at (700,-2) {$\phi_e$ = 0.666};
	\addlegendentry{$\varepsilon = 0.40$}
\end{axis}
\end{tikzpicture}%
~%
	\begin{tikzpicture}
 
	\begin{axis}[
    axis lines = left,
    xlabel = $\phi$,
	ylabel = {$\rho(\phi)$},    
     ymin=0, ymax=1,
    legend pos = north west,
     ymajorgrids=true,
    xmajorgrids=true,
    grid style=dashed,
    width=5cm,
    height=5cm,
     ]

\addplot [
dashed,	
	domain= 0:1,
	color= blue,
	]
	{(0.1*x)/((0.1*x+(1-0.15)*(1-x))}; 
	\node[above,red] at (750,-2) {$\phi_e$ = 0.74};
	\addlegendentry{$\varepsilon = 0.20$}
\end{axis}
\end{tikzpicture}
	
\newpage
\subsection{Example of SARS-CoV-2 Pandemic}

The current COVID-19 pandemic provides an excellent opportunity to apply the methods herein described.
The nasal swab PCR screening test for COVID-19 has been shown to have a high analytical sensitivity of 95 percent limit of detection (LOD) for the RNA-dependent RNA polymerase (RdRP) gene. Likewise, the test is 99 percent specific for SARS-CoV-2 when tested against 31 common respiratory pathogens \cite{cheng2020diagnostic}. We thus draw the screening curve for this test $\rho(\phi)$ (Figure 4):
\\
\

\begin{center}

\begin{tikzpicture}
 
	\begin{axis}[
    axis lines = left,
    xlabel = $\phi$,
	ylabel = {$\rho(\phi)$},    
     ymin=0, ymax=1,
    legend pos = south east,
     ymajorgrids=true,
     xmajorgrids=true,
    grid style=dashed,
    width=8 cm,
    height=8cm,
     ]
	\addplot [
	domain= 0:1,
	color= blue,
	]
	{(0.95*x)/((0.95*x+(1-0.99)*(1-x))};
	\node[above,red] at (850,90) {$\phi_e$ = 0.093};
	\addlegendentry{$\varepsilon = 1.94$}
	\end{axis} 
	 \draw [dashed, red, thick] (0.56,0) -- (0.56,5.76);
\end{tikzpicture}
\end{center}
\textbf{Figure 4.} The screening curve for the SARS-CoV-2 nasal PCR test (blue) and the prevalence threshold level (red) below which the positive predictive value drops.
\\
\

We calculate the prevalence threshold $\phi_e$ by using equation (28), with values for a = 0.95, b = 0.99 and therefore $\varepsilon$ = 1.94. We thus obtain:

\begin{large}
\begin{equation}
\phi_e =\frac{\sqrt{a\left(-b+1\right)}+b-1}{(\varepsilon-1)} =\frac{\sqrt{.95\left(-.99+1\right)}+.99-1}{(1.94-1)} = .093
\end{equation}
\end{large}

As noted in the figure above, significant drops in prevalence only marginally impact the PPV until the prevalence threshold is reached. In other words, when the prevalence of active COVID-19 cases drops below 9.3 percent, the nasal RT-PCR test's PPV drops significantly faster. Since 9.3 percent of the population has thankfully not been infected at any given time, we deduce that a significant proportion of the current positive nasal RT-PCR tests are false positives. The benefits of contextualizing the validity of a screening test in real time cannot be understated This is indeed a critical exercise since a large number of public health decisions rely on the validity of these screening tests. With a reliable test, we can better inform the individual on his or her risk of contracting and transmitting the disease in question. Likewise, it can guide quarantine guidelines so as to best integrate that individual back into the economy and society at large. Furthermore, reliable estimates of incidence and prevalence with good tests can guide the proper distribution of resources to contain the spread of the virus. All in all, understanding where this prevalence point lies in the curve has important implications for the administration of healthcare systems, the implementation of public health measures, the development of epidemiologic models, and in cases of pandemics like SARS-CoV-2, the functioning of society at large. When the prevalence drops below the prevalence threshold, the censoring of patients never affected needs to be contrasted with the Bayesian limitations imposed by the screening paradox.

\section{Conclusion}
The curvilinear relationship between a screening test’s positive predictive value and its target disease prevalence is proportional. In consequence, there is an inflection point of maximum curvature in the screening curve defined as a function of the sensitivity and specificity beyond which the rate of change of a test's PPV declines sharply relative to disease prevalence. Herein, we demonstrate a mathematical model exploring this phenomenon and define the prevalence threshold point where this change occurs. To the best of our knowledge, while this concept is a simple consequence of Bayes' theorem and the natural shape of screening curves, it has never been properly formalized mathematically as showcased in this work. The prevalence threshold can help contextualize the validity of a screening test in real time, thereby enhancing our understanding of the dynamics and epidemiology of specific conditions. Finally, this simple equation can be applied to any and all screening test whose sensitivity, specificity and target prevalence are known - so its clinical utility is widespread. 

\newpage

\bibliographystyle{unsrt}
\bibliography{references}

\begin{thebibliography}{1}

\bibitem{wilson1968principles}
James Maxwell~Glover Wilson, Gunnar Jungner, World~Health Organization, et~al.
\newblock Principles and practice of screening for disease.
\newblock 1968.

\bibitem{sackett1975screening}
DAVID~L Sackett.
\newblock Screening for early detection of disease: to what purpose?
\newblock {\em Bulletin of the New York Academy of Medicine}, 51(1):39, 1975.

\bibitem{andermann2008public}
Anne Andermann, Ingeborg Blancquaert, Sylvie Beauchampb, and V{\'e}ronique
  D{\'e}ryc.
\newblock Public health classics.
\newblock {\em Bulletin of the World Health Organization}, 86(4), 2008.

\bibitem{herman2006makes}
Cheryl Herman.
\newblock What makes a screening exam" good"?
\newblock {\em AMA Journal of Ethics}, 8(1):34--37, 2006.

\bibitem{brenner1997variation}
Hermann Brenner and OLAF Gefeller.
\newblock Variation of sensitivity, specificity, likelihood ratios and
  predictive values with disease prevalence.
\newblock {\em Statistics in medicine}, 16(9):981--991, 1997.

\bibitem{moons1997limitations}
Karel~GM Moons, Gerrit-Anne van Es, Jaap~W Deckers, J~Dik~F Habbema, and
  Diederick~E Grobbee.
\newblock Limitations of sensitivity, specificity, likelihood ratio, and bayes'
  theorem in assessing diagnostic probabilities: a clinical example.
\newblock {\em Epidemiology}, pages 12--17, 1997.

\bibitem{fawcett2006introduction}
Tom Fawcett.
\newblock An introduction to roc analysis.
\newblock {\em Pattern recognition letters}, 27(8):861--874, 2006.

\bibitem{smith2000first}
James~E Smith, Robert~L Winkler, and Dennis~G Fryback.
\newblock The first positive: computing positive predictive value at the
  extremes.
\newblock {\em Annals of internal medicine}, 132(10):804--809, 2000.

\bibitem{cheng2020diagnostic}
Matthew~P Cheng, Jesse Papenburg, Micha{\"e}l Desjardins, Sanjat Kanjilal,
  Caroline Quach, Michael Libman, Sabine Dittrich, and Cedric~P Yansouni.
\newblock Diagnostic testing for severe acute respiratory syndrome--related
  coronavirus-2: A narrative review.
\newblock {\em Annals of internal medicine}, 2020.

\end{thebibliography}

\end{document}